\documentclass[journal,twoside,web]{ieeecolor}
\usepackage{jsen}
\usepackage{cite}
\usepackage{amsmath,amssymb,amsfonts}
\usepackage{algorithmic}
\usepackage{graphicx}
\usepackage{textcomp}
\usepackage{wrapfig}
\usepackage{comment}
\usepackage{booktabs}
\usepackage{subcaption}
\usepackage{siunitx}
\usepackage{hyperref}

\def\BibTeX{{\rm B\kern-.05em{\sc i\kern-.025em b}\kern-.08em
    T\kern-.1667em\lower.7ex\hbox{E}\kern-.125emX}}
\definecolor{abstractbg}{rgb}{0.89804,0.94510,0.83137}
\newcommand{\rebu}[1]{\textcolor{black}{#1}}

\definecolor{gray}{rgb}{0.4, 0.4, 0.4}

\usepackage{eso-pic}
\usepackage{url}
\AddToShipoutPictureBG*{
  \AtPageUpperLeft{%
    \put(0,-40){\raisebox{18pt}{\makebox[\paperwidth]{\begin{minipage}{21cm}\centering
      \textcolor{gray}{This article has been accepted for publication in the IEEE Sensors Journal (JSEN). \\ DOI: 10.1109/JSEN.2024.3424655}
    \end{minipage}}}}%
  }
  \AtPageLowerLeft{%
    \raisebox{20pt}{\makebox[\paperwidth]{\begin{minipage}{21cm}\centering
      \textcolor{gray}{ \copyright 2024 IEEE.  Personal use of this material is permitted.  Permission from IEEE must be obtained for all other uses, in any current or future media, including reprinting/republishing this material for advertising or promotional purposes, creating new collective works, for resale or redistribution to servers or lists, or reuse of any copyrighted component of this work in other works.
      }
    \end{minipage}}}%
  }
}

\begin{document}

\title{\textbf{S}ep\textbf{Al}: \textbf{S}epsis \textbf{Al}erts On Low Power Wearables With Digital Biomarkers and On-Device Tiny Machine Learning}
\author{Marco Giordano \IEEEmembership{Student, IEEE}, Kanika Dheman \IEEEmembership{Member, IEEE}, and Michele Magno \IEEEmembership{Senior, IEEE}\\}

\renewcommand{\arraystretch}{1.5}

\IEEEtitleabstractindextext{%
\fcolorbox{abstractbg}{abstractbg}{%
\begin{minipage}{\textwidth}%
\begin{wrapfigure}[12]{r}{3.4in}%
\includegraphics[width=3.2in,trim=0cm 4cm 0cm 0.7cm,clip]{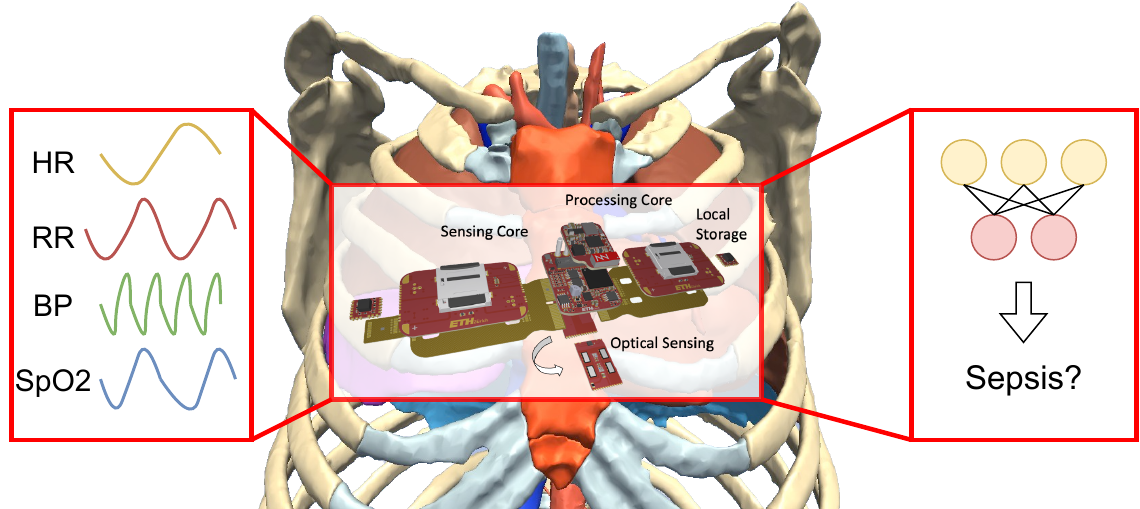}%
\end{wrapfigure}%
\begin{abstract}
Sepsis is a lethal syndrome of organ dysfunction that is triggered by an infection and claims 11 million lives per year globally. 
Prognostic algorithms based on deep learning have shown promise in detecting the onset of sepsis hours before the actual event but use a large number of bio-markers, including vital signs and laboratory tests. The latter makes the deployment of such systems outside hospitals or in resource-limited environments extremely challenging. This paper introduces SepAl, an energy-efficient and lightweight neural network, using only data from low-power wearable sensors, such as photoplethysmography (PPG), inertial measurement units (IMU), and body temperature sensors, designed to deliver alerts in real-time. SepAl leverages only six digitally acquirable vital signs and tiny machine learning algorithms, enabling on-device real-time sepsis prediction.  

SepAl uses a lightweight temporal convolution neural network capable of providing sepsis alerts with a median predicted time to sepsis of 9.8 hours. The model has been fully quantized, being able to be deployed on any low-power processors, and evaluated on an ARM Cortex-M33 core. Experimental evaluations show an inference efficiency of 0.11MAC/Cycle and a latency of 143ms, with an energy per inference of 2.68mJ. This work aims at paving the way toward accurate disease prediction, deployable in a long-lasting multi-vital sign wearable device, suitable for providing sepsis onset alerts at the point of care.

The code used in this work has been open-sourced and is available at \href{https://github.com/mgiordy/sepsis-prediction}{https://github.com/mgiordy/sepsis-prediction}.
\end{abstract}

\begin{IEEEkeywords}
Sepsis Onset, Edge Computing, Tiny Machine Learning, Wearable devices, Smart Sensors
 \underline
{}
\end{IEEEkeywords}
\end{minipage}}}

\maketitle

\section{Introduction}
\label{sec:introduction}
Sepsis is a syndrome of organ dysregulation that is triggered by an infection. Sepsis has a worldwide incidence of 48.9 million cases and accounts for about 11 million deaths ( approximately 20$\%$ of all global deaths) \cite{wentowski2021sepsis}.
Identifying the onset of sepsis is challenging because of the heterogeneous clinical presentations of sepsis in people of different demographics and the presence of co-morbidities \cite{fernando2018prognostic}. Early recognition and rapid initiation of antibiotics and supportive management are critical in sepsis as no specific therapy is available to clinicians\cite{johnston2017effect}. It has been shown that mortality linearly increases by 0.42$\%$ per hour of delay\cite{Ruddel2022} to administer antibiotics in sepsis patients. For patients with septic shock, a delay in antibiotic administration increased the risk of mortality by 35$\%$ \cite{Im2022}. \\

\begin{table*}
\centering
  \begin{tabular}{cccccccccc}
    \toprule
    Reference & Dataset & Labelling & Model & Parameter$^a$ & Input Size & \rebu{Retro/} & KPIs \\
     &  &  &  &  &  & \rebu{Online} &  \\
    \midrule
        \cite{nemati2018interpretable} & Emory & Sepsis-3 & Weibull-Cox & 65 features & all data & \rebu{R} &  Pred$^e$: 4h, AUROC: 0.85, Spec$^h$: 0.67 \\
    \cite{henry2015targeted} &MIMIC-II & ICD-9 codes & Cox-Prop. & 54 features &all data & \rebu{O} & Pred$^f$: 28.2h, Sens$^i$: 0.85, Spec$^h$: 0.67 \\
    \cite{liu2022dynamic}& MIMIC-III,PC$^b$ & Sepsis-3 & XGBoost & 63 Features & all data & \rebu{R} & AUROC: 0.82, F1: 0.165 \\
    \cite{rapid_response} & CNUH,UV$^c$ & Sepsis-3 & Graph Atten. & 7 V; 24 LT & 8-24 h & \rebu{O} & At onset$^g$, AUROC: 0.93, AUPRC: 0.86 \\
    \cite{multi_branch}& PC$^b$  & Sepsis-3 &TCN & 8 V; 26 LT; 6 D & all data & \rebu{R} & Pred$^e$: 6h, AUROC: 0.89, AUPRC: 0.52 \\
    \cite{moor2019early}& MIMIC-III & Sepsis-3 & TCN & 44 V and LT & all data  & \rebu{R} & Pred$^e$: 7h, AUROC: 0.87, AUPRC: 0.42 \\
    \cite{Mao2018}& Multi-center$^d$ & SIRS& GBT & 6V & all data & \rebu{R} & At onset, AUROC: 0.74, AUPRC: 0.28 \\
    \cite{Ding_2023} & MIMIC-III & Sepsis-3 & Semi-superv. & 7 V; 21 LT; 4 D& 6 h   & \rebu{R} & Pred$^e$: 6h, AUROC: 0.76 \\
  \bottomrule
\end{tabular}

\scriptsize
[a]  V: Vitals; LT: Lab Tests; D: Demographic data. [b] Physionet Challenge. [c]  CNUH: Chonnam National University Hospital; UV: University of Virginia. [d]  Mixed ward, multi-center dataset from the University of California, San Francisco (UCSF), Medical Center (San Francisco, California), and Beth Israel Deaconess Medical Center (Boston, Massachusetts); [e] Prediction time; [f] Median prediction time to sepsis shock; [g]  After multiple abnormal events. [h] Specificity; [i] Sensitivity.

\caption{State of the art in predicting sepsis onset}
\label{tab:relatedwork}
\end{table*}

At present, sepsis monitoring is done in clinical environments with medical scoring systems that track changes in vital signs and laboratory-obtained biomarkers that are indicative of organ dysfunction.
Sepsis- 3 defined a Sequential Organ Failure Assessment (SOFA) score which was based on multiple laboratory tests and vitals specific to each organ system, such as the respiratory, cardiovascular, renal, liver, central nervous system, and blood coagulation.
In addition, the modified early warning score (MEWS), quick SOFA (qSOFA), or APACHE II  have also been used for sepsis identification, patient health deterioration, or severity of disease classification. However, all these scores are punctual and require iterative measurements from staff, which becomes more sparse when the patient is outside the \rebu{Intensive Care Unit (ICU)}, be it a general ward or at home.

Vital sign measurement is one of the key components of calculating the aforementioned scores. Clinical practice at present relies on manual sporadic measurement by clinical staff outside the ICU and using patient monitors in the ICU. This is a major drawback because of (a) the need for dedicated resources, infrastructure, and maintenance, (b) the availability of continuous vital sign monitoring restricted to the intensive care units (ICUs), and (c) the connection of the monitors to the patient via multiple wires making movement difficult and resulting in injury or skin lacerations in specific groups such as the pediatric or neonatal population. In addition, the delay in transferring the manually recorded data to the electronic database causes high false alarms and subsequent alarm fatigue when used in predictive disease modelling\cite{schwab2018wolf}.

Wearable vital sign monitoring can automate the process of generating and updating electronic medical records (EMR) at a high temporal density. Delays in EMR reporting can be eliminated with autonomous and low-power multi-sensor wearable systems based on the always-on smart sensing paradigm where physiological data can be continuously and comfortably streamed or processed to extract vitals in real-time in a sensor node. To achieve this goal, the wearable system must rely solely on measurable vital signs from low-power, compact sensors like IMUs and PPGs \cite{gupta2022higher,magno2019self}.
Deep learning approaches have shown promising results in predicting sepsis early with models trained on open source datasets from ICU stays where the time to predict sepsis onset typically ranges from 4 to 24 hours\cite{Moor2021}. However, these models are not suitable for on-device processing on wearable devices and they require GPU clusters for inference mostly due to their large input feature size that comprises sparse data from sporadic lab tests and low granularity vital sign measurements. Moreover, previous works use databases with information not always available on wearable devices.  Having wearable devices with on-device intelligence enables the implementation of such tools outside the ICU, especially in low and middle-income countries (LMIC). 

Deep learning is often associated with high computational and memory requirement and require infrastructure with GPUs to run. Recent advancements in low-power processing units, such as MCUs, and efficient machine learning offer promising opportunities to bring machine-learning models on wearable devices, combining sensing and computing in the same sensor node. Tiny Machine Learning (TinyML) has been gaining traction in the low-power embedded systems\cite{9969601,10190347} and machine learning community\cite{9912325} and has been showing promising results in the biomedical field\cite{10236992,9743469}. 
TinyML is enabling predictive disease modeling at the point of care, however, several challenges remain open in this perspective. For instance, a consequence of using (almost) all of the EMR results in a high dimensional input vector. This renders existing deep neural network models unsuitable for low-power, battery-operated devices capable of continuous patient monitoring \cite{Moor2021}, which are limited in terms of memory and computational capabilities. 

This paper presents SepAl, an energy-efficient intelligent algorithm targeted to low-power wearable devices. Leveraging only 6 vital signs, that can be extracted by IMUs and PPG, which can be embedded in wearable devices and do not require laboratory infrastructure, its goal is to bring the detection of sepsis onset at the edge.
This approach is particularly challenging though, since many diseases have a direct correlation with physical quantities captured only by laboratory tests and more complex dependence on vital sign time trends. Thus, this paper investigates the possibility of measuring only 6 vital signs as inputs of a developed quantized lightweight temporal convolution neural network, which can continuously run on a wearable device.
Bringing data processing close to the sensors on wearable devices reduces the delay in data gathering, improves energy efficiency, reduces the latency of the detection, and improves the quality of care without the need for dedicated infrastructure or impinging upon the privacy of the patient.

The major contributions of this work are:
\begin{itemize}
     \item Development and evaluation of a temporal convolution network (TCN) that uses only digital biomarkers as inputs for detecting the onset of sepsis in realistic real-time conditions using low-power processors such as  Arm Cortex-M33 cores. 
     \item Investigation and evaluation of the sepsis prediction using only 6 vital signs that can be extracted by commercially available IMUs and PPG sensors.
     \item Experimental evaluation of the developed algorithm in terms of sensitivity, specificity, energy, latency, and time to sepsis prediction. 
 \end{itemize}

The code used in this work has been open-sourced and is available at \href{https://github.com/mgiordy/sepsis-prediction}{https://github.com/mgiordy/sepsis-prediction}.
\section{Related works}

Various methods based on both statistical and deep learning techniques have been used before to provide a data-driven prediction of sepsis, as summarised in Table \ref{tab:relatedwork}.  A major challenge in modeling the probability that a patient might develop sepsis is a result of the severe class imbalance for sepsis in open-source EMR databases as reported in the literature for various datasets\cite{multi_branch}\cite{Moor2021}. The low incidence of sepsis results in a high number of false positives leading to alarm fatigue. In addition, open-source electronic medical records often have low granularity data, missing values, and erroneous values (when taken from handwritten notes).

\begin{figure*}
    \centering
    \begin{minipage}{0.5\textwidth}
        \centering
        \includegraphics[width=0.8\textwidth]{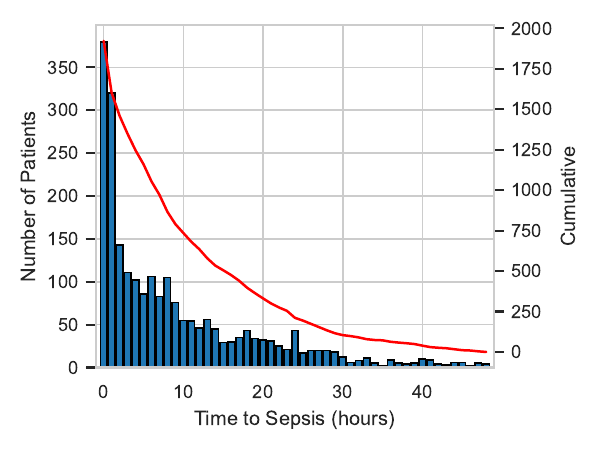}
        \subcaption{}
    \end{minipage}
    \begin{minipage}{0.49\textwidth}
        \centering
        \includegraphics[width=0.9\textwidth,trim=0 0.6cm 0 0,clip]{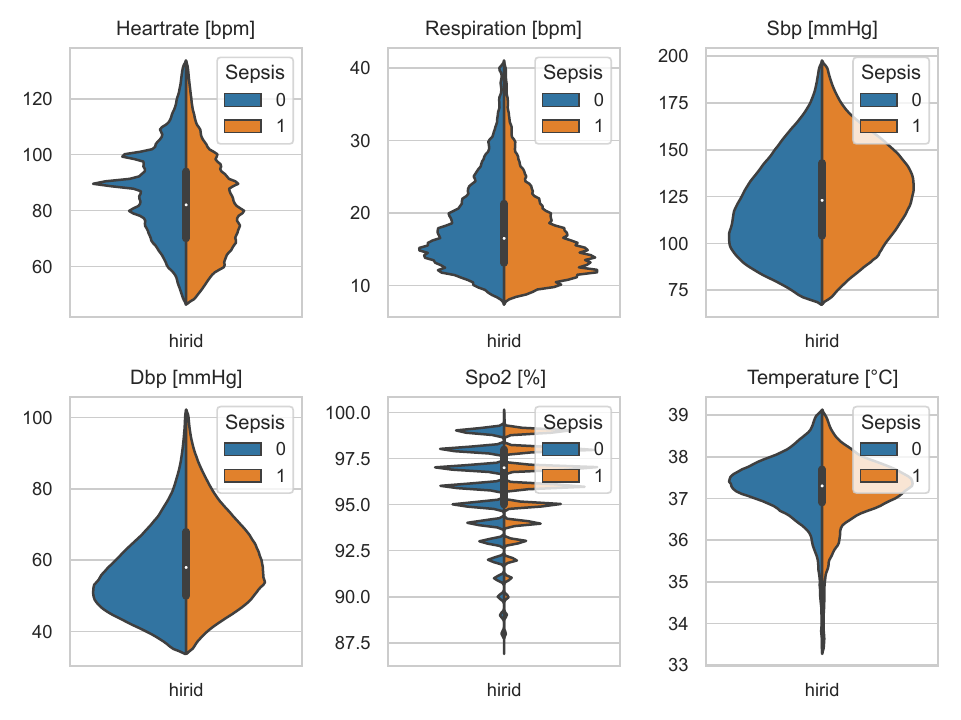}
        \subcaption{}
    \end{minipage}
   \caption{(a) Histogram of sepsis-positive patients in the HiRID dataset. (b) Violin plots of the distributions of the vital signs considered in this work: heart rate, respiration rate, systolic blood pressure (sbp), diastolic blood pressure (dbp), SpO2, and body temperature.}
   \label{fig:dataset_stats}
\end{figure*}

Sepsis prediction using Weilbull-Cox proportional hazards modeled 65 features extracted from the Emory University hospitals dataset and MIMIC-III with a sepsis incidence of 8.6\% \cite{nemati2018interpretable}.
The model AUROC ranged from 0.85 to 0.83  for prediction horizons of 4h to 12h with a reported sensitivity of 0.85 and specificity of 0.67. Similarly, TREWScore-based on the Cox proportionality hazard model used 54 features to predict the onset of septic shock and analyzed the data per hour to calculate a risk score \cite{henry2015targeted}. They reported a median prediction horizon of 28.2 hours before the onset of septic shock with a sensitivity of 0.85 and specificity of 0.67. However, they focused only on septic shock and organ failure, targeted at the further deterioration of a patient already diagnosed with sepsis and without the ability to pre-empt the likelihood of contracting sepsis right at the onset of infection.
However, it must be noted that the TREWScore was tested rigorously in the ICU with a 2-year study that showed a decrease in mortality due to pre-emptive artificial intelligence solutions \cite{Adams2022},\cite{Henry2022}. However, such a highly parameterized model could be deployed only in the ICU of a clinical facility with the necessary infrastructure, that is, the availability of automated EMRs and GPUs for processing and laboratories capable of running biological tests with a fast turnaround.\\

A limitation of previous works dealing with sepsis onset prediction is that their analysis is performed retrospectively, i.e. classifying already occurred events, and therefore not implementable in a real-time online device. An XGBoost model was implemented for a retrospective task of identifying sepsis-positive patients with 63 handcrafted features and using all the variables available in the EMR dataset \cite{liu2022dynamic}. It reported an AUROC of 0.82 and a weighted F1-score of 0.82. While this model performed very well with respect to the reduction in the false positives, its suitability for deployment for edge computing is still limited due to the retrospective mode of analysis and dependence of the large parametric model on all EMR variables employed in feature engineering.

A double graph attention network was used to model multivariate correlations retrospectively between 31 input parameters to detect abnormal changes in patients as anomalies \cite{rapid_response} and achieved an AUROC of 0.93 and AUPRC of 0.86. In another approach, a temporal convolution network (TCN) applied a k-nn-based model with dynamic time warping that leveraged data sparsity by interpreting the missing data with a Gaussian process \cite{moor2019early}. The model used 44 irregularly sampled laboratory and vital parameters for predicting the onset of sepsis 0 to 7h preceding sepsis onset and provided an AUROC ranging from 0.9 to 0.87 and an AUPRC ranging from 0.5 to 0.42, respectively. In another implementation of the TCN, the authors use all available input variables in the dataset and address the value of missing-ness by masking the missing data in a parallel branch of the neural network instead of performing forward or back imputation \cite{multi_branch}. Moreover, they address data imbalance with a k-nn-based undersampling approach. The authors report an AUROC of 0.892 and AUPRC of 0.527 for a prediction horizon of 6h.\\

On the other hand, the \textit{InSight} model based on gradient tree boosting using only 6 vitals signs has also been vigorously tested with multiple research, economic, and clinical trials \cite{Calvert2016},\cite{Desautels2016},\cite{Shimabukuro2017}. In a multicenter, multi-ward setting validation of sepsis prediction with a large dataset of 90,353 patients resulted in a high AUROC reported by the authors and reduced false positives\cite{Mao2018}.
However, while the authors provided a prediction of shock at a horizon of 4 hours, they provided the sepsis onset and severe sepsis assessment at the onset of the events and not before. Other than still being a retrospective analysis, this does not provide a pre-emption of the event and misses the main target of deploying such models.\\

\rebu{
To bring related work into perspective, 5 out of the 8 surveyed papers used the open source datasets MIMIC or Physionet Challenge, which have a limited data granularity (1 hour sampling period). We used the HiRID dataset, collected at the Bern University Hospital and featuring a high granularity (2 minutes sampling period).
Our work, along with 6 out of the 8 surveyed papers, used Sepsis-III\cite{singer2016third} as the definition of Sepsis, which is the most comprehensive to date.
A core contribution of this paper is to demonstrate that it is possible to detect Sepsis early with only digital biomarkers. All the papers surveyed have a high number of features, apart from \cite{Mao2018}, which uses the same six input variables as ours. However, \cite{Mao2018} considered the problem only in a retrospective manner and targeted the prediction only at the onset of Sepsis. In our work, we provide results both as a retrospective task, with a prediction horizon of 4 hours, and a more challenging, but real-time, online task.
}

In this work, we have addressed the challenge of sepsis detection by developing a deep learning algorithm that can run on a wearable device only with vital sign information extracted by IMUs, a PPG sensor, and a body temperature sensor.
Unlike previous works that proposed large, memory-intensive, complex algorithms, and retrospective analysis, this work introduces a lightweight algorithm utilizing only low-power sensors and low-power processors enabling a truly intelligence wearable system.
The presented work aims to provide an accurate evaluation of the proposed approach. Such a system would be ubiquitous, provide automated integration of continuously monitored vital signs, and protect patient privacy while giving access to high granularity and diverse data for adverse event modeling and pre-emption. By leveraging these advancements, we aim to enhance the feasibility and practicality of wearable devices in sepsis monitoring, ultimately improving patient outcomes.
\begin{figure*}[hbt!]
    \centering
\includegraphics[width=0.9\textwidth]{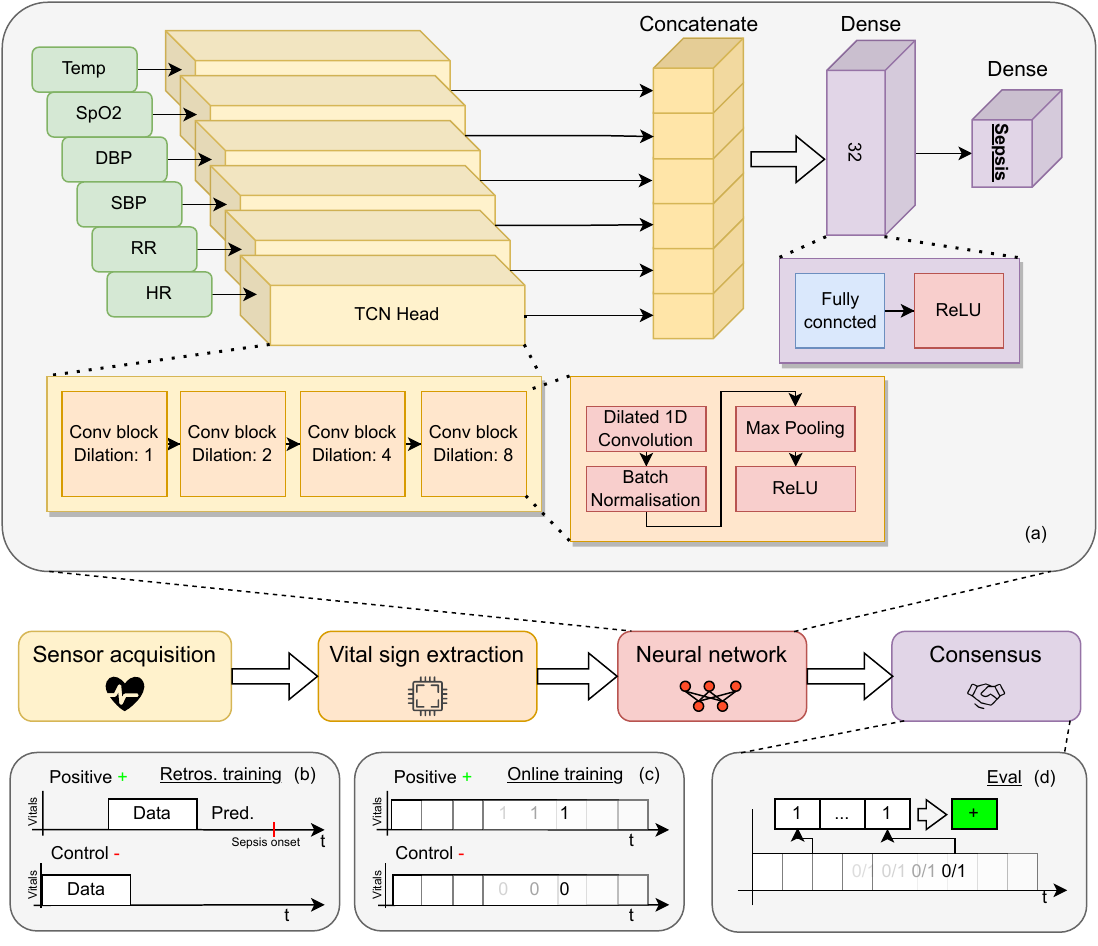}
    \caption{Data processing pipeline from sensor data acquisition to on-board vital sign extraction, which consists of multiple digital signal processing steps. After vital sign extraction, the vitals are fed to the neural network and then to the consensus algorithm for the prediction of sepsis onset. (a) Multi-modal temporal convolution neural network architecture. (b) Data windows organization for the retrospective analysis. (c) Data windows organization and labeling for neural network real-time training. (c) Data windows prediction and consensus algorithm used for model evaluation and validation in real-time.}
    \label{fig:vital_extraction}
\end{figure*}

\section{System Architecture}

\label{section: Model_TCN_arch}
Predictive modeling to identify sepsis is done at present using all the parameters of the EDMs. However, ubiquitous monitoring and detection of sepsis at the point of care on wearable devices can only be made feasible if sepsis onset can be identified using digital biomarkers of health, such as vital signs. Hence, to evaluate if it is even feasible to identify sepsis onset using just the vital signs, a model based on only digitally acquired vital signs from the EMR  was developed to identify sepsis patients retrospectively. Finally, a real-time sepsis onset detection model was trained and evaluated to identify sepsis temporally at a chosen stride.

\subsection{Multi-head temporal convolution network (TCN)}

A multi-head TCN was developed to achieve an accurate and lightweight algorithm for continuous, always-on analysis targeting implementation on low-power microcontrollers. Moreover, TCNs are designed to determine causality in time series data which makes them appropriate for use in identifying sepsis in real-time \cite{bai2018empirical,9765481}.
The architecture of the TCN is shown in Figure \ref{fig:vital_extraction}.  Previous work employing TCN architectures for sepsis detection \cite{moor2019early}, \cite{multi_branch},\cite{Rosnati2021}, use a single TCN model that learned multivariate correlations between data streams passed collectively to the TCN. The model proposed in this work leverages a singular TCN per input variable to independently learn the causal dependence of each input. The outputs of all the \emph{n} TCNs are then combined by concatenating and flattening into a dense layer. This design choice favors the information fusion in the dense layers instead of the convolutions layers such that the model can learn individualistic features from each input stream.\\ 

 The input shape of the proposed TCN is $\text{vital signs} \times \text{data points}$, which in the context of this analysis is $\text{n}\times\text{4h * 2 samples/min} = \text{n}\times\text{120}$, forming a matrix with one vital sign per row and the columns being the corresponding time series. Data have been normalized by subtracting the mean and dividing by the standard deviation of the training dataset, with the same normalization values being also applied to the test set.
 
 The model hyper-parameters, like the numbers of layers, convolution filters, learning rate, and schedule were optimized with a weight-informed Neural Architecture Search (NAS)\cite{ren2021comprehensive}. NAS samples random combinations from a given search space and fits models to defined memory constraints after training and evaluation. The architecture search yielded the best model to have 4 layers of dilated causal convolution, with a power-of-2 exponential increase in dilation size to capture longer-range dependencies in the time-series data and with 32 filters each. The output of the convolution layer is then flattened and fed into a dense layer of 32 neurons. The kernel dimension of each convolution layer is set to 3. Batch normalization is applied after each convolution layer followed by a max pooling layer with kernel and stride of 2 and a ReLU activation function. The last layer, instead of a ReLU activation function, has a Sigmoid function, allowing for an interpretation of the output as a probability.

\subsection{Input data vector}
The input of the model consists of 4 hours of data sampled at 2 samples/minute using the high-granularity HiRID dataset. Digital biomarkers from the EMR that were considered while developing the model are the heart rate (HR), systolic blood pressure (SBP), diastolic blood pressure (DBP), respiratory rate (RR), core body temperature, and peripheral blood oxygen level (SpO2). 

\rebu{
The selection is motivated by the feasibility of acquiring these markers with wearable sensors, making the system non-dependant on external laboratories and/or manual data entries. In particular, HR, RR, and SpO2 are obtainable from photoplethysmograph (PPG). PPG sensors are present in several smart-watches, smart bands, and smartphones, and are an object of recent efforts in academia\cite{chen2020modulation,naeini2019real}. PPG sensors are relatively cheap, and their power consumption has been optimized to run on wearable battery-powered devices: recent sensors show average active power in the order of a few milliwatts\footnote{https://www.analog.com/en/products/max86178.html}. Body temperature can be measured by specialized sensors, which are also increasingly inexpensive and show excellent low-power performance, with just a few milliwatts required in active mode\footnote{https://www.analog.com/en/products/max30208.html}. Lastly, blood pressure is measurable by accelerometers\cite{chang2019cuff,das2021noninvasive,hsu2021motion} fulfilling the AAMI criteria. Accelerometers are sensors nearly omnipresent in wearable devices, with low cost and excellent low-power performance, consuming as low as tens of \qty{}{\micro\watt} while constantly sampling\footnote{https://www.st.com/en/mems-and-sensors/iis2dlpc.html}.
}

\rebu{The model proposed in this work needs a reading of vital signs every 2 minutes. This means that the sensors needed to acquire such vitals can be duty-cycled to further lower their power consumption. In particular, a PPG sensor usually takes a few seconds of active mode to acquire a signal from which HR, RR, and SpO2 can be extracted. Similarly, temperature sensors usually take only a few tens of milliseconds per acquisition, after which they can enter sub-mW sleep modes. Accelerometers, on the other hand, must continuously collect data, usually with sampling frequencies of a few hundred hertz. However, accelerometers have been optimized for such an operation, managing to stay in the sub-milliwatt range while periodically sampling.}

\rebu{
Summarising, the vital signs selected can be acquired with inexpensive and low-power sensors. This enables SepAl to be evaluated on highly integrated and inexpensive hardware, with expected long-lasting battery lifetimes.
}

\section{Methods}
\subsection{Dataset} The HiRID dataset has 34 thousand patient records from the Department of Intensive Care Medicine of the Bern University Hospital, Switzerland. It provides unidentified demographic data, real-time measurements from bedside monitors,  usage of medical devices (i.e. mechanical ventilation), observation notes by health care providers, laboratory-acquired biochemical markers, administered drugs, fluids, and nutrition\cite{yeche2021hirid}. 
To manage the high number of records in the database, the data is loaded into a Postgres instance running on an on-premise server. A data loader downloads the patient list and then accesses the database generating SQL queries to fetch the user data, process the data to apply necessary inclusion criteria, labels, and imputations, and then save the data in CSVs files, which are used as input to the neural network.

Figure \ref{fig:dataset_stats}(a) plots the distribution of sepsis-positive patients across the time to sepsis onset. As shown by the histogram, more than half of the total sepsis-positive patients develop sepsis within the first 8 hours of ICU stay. This limits the amount of data available from training machine learning models and worsens the class imbalance. Figure \ref{fig:dataset_stats}(b) shows the distribution plots of the vital signs in the EMR that can be digitally acquired via wearable devices for both sepsis-positive and negative patients. Some vital signs, like blood pressure and heart rate, show a significant skew in the distribution between the two cohorts.

\subsection{Patient selection and sepsis onset labeling} Patient selection was based on inclusion criteria to have a patient cohort consistent with previous works. This included having a length of stay of at least 24 hours, a minimum age of 18 years, exclusion of patients administered antibiotics in the first 7 hours of ICU admission, and exclusion of patients with time to sepsis onset of less than 4 hours (as the length of at least 4 hours was chosen as the input data window). The maximum length of ICU stay was terminated at 48 hours to balance the positive and negative windows during training. The implementation of this criteria resulted in 1058 sepsis-positive patients and 7635 sepsis-negative patients and a class imbalance of 13.9$\%$. The class imbalance was handled by under-sampling the highest represented class (sepsis-negative) of control cases to achieve an equal number of positive and negative cases.
The data was split into training and test sets with an 80/20 ratio respectively. To avoid altering the test data in any way, and to have a fair comparison for real-life conditions, only the train set was balanced, while the test set did not contain any under-sampled or clipped data.

The time for sepsis onset was calculated and annotated according to the sepsis-3 \cite{singer2016third} criteria. A window for suspicion of infection is defined by first identifying the time point where antibiotics are administered and then taking a time window 48 hours before and 24 hours ahead. The time for sepsis-onset is marked for a patient at the point when the hourly difference of SOFA score is greater or equal to 2. For the retrospective identification of sepsis patients, 4 hours windows were extracted at a varying time before the Sepsis onset for sepsis-positive patients, and at the beginning of ICU stay for controls, as shown in Figure \ref{fig:vital_extraction}(b). The entire window for a sepsis-positive patient was labeled as class 1 while a sepsis-negative patient was labelled as class 0. To train the real-time sepsis onset identification, an assumption was made that the features specific to sepsis positive are present right from the beginning, and hence, each data window of 4 hours was labeled as a class 1 for the sepsis positive and a class 0 for all negative patients, Figure \ref{fig:vital_extraction}(c). The consensus algorithm applied on successive windows for the real-time evaluation is shown in Figure \ref{fig:vital_extraction}(d).

\subsection{Model training and evaluation}
\label{sec:dl}

\subsubsection{Retrospective sepsis patient identification}

As explained in section \ref{section: Model_TCN_arch}, the designed TCN model was first trained to identify septic patients retrospectively to establish whether only digitally acquired vital signs could identify patients of sepsis. In order to do this, a prediction window was marked backward from the sepsis onset time. An input data window with a span of 4 hours was chosen at the end of the prediction window for each of the 6 vital signs. The model was trained to predict a septic patient with a prediction horizon of 1 hour, 2 hours, 3 hours, and 4 hours in the future. \\ 

\subsubsection{Real-time sepsis onset identification}

A realistic temporal model for an edge device was trained to give a real-time prediction of sepsis onset by splitting the data in rolling windows with a given stride. For every 4-hour data window, a prediction was made by the TCN model to detect whether a certain window is sepsis-positive or negative. Thereafter, an aggregation of the positive predicted class was made to raise an alarm for sepsis as shown in Figure~\ref{fig:vital_extraction}. This consensus strategy dictated that if \textsc{k} consecutive windows have been labeled as positive, then the patient is classified as sepsis positive. This way the first prediction by the algorithm would be at: 
\begin{equation*}
\text{size(input data window)} + \textsc{k} \cdot \text{window stride}
\end{equation*}
The time of the last window is considered to be the prediction time for sepsis onset. To optimize this algorithm, the stride of the data window was parameterized with lengths of 10 minutes, 30 minutes, and 60 minutes. These values were chosen so to consider data window overlaps of different extents and also have reasonable time to first prediction. In addition, the aggregation parameter \textsc{k} was also changed from 2 to 12 to evaluate the prediction sensitivity and specificity.\\

The TCN for both the two methodologies described above has been trained with a step learning rate that was multiplied by 0.2 every 4 epochs, for a total of 10 epochs. Adam\cite{kingma2014adam} has been chosen as the optimizer and Binary Cross Entropy as the loss function.
\\
\subsubsection{Evaluation metrics}
\label{sec:eval}
The models for both training cases: retrospective septic patient identification and real-time sepsis onset identification, were primarily evaluated using metrics of sensitivity and specificity to account for false alerts. In addition, for the real-time identification of sepsis onset, the median predicted time to sepsis onset was calculated from different window strides and aggregation parameter \textsc{k} as a marker of the predictive power of the algorithm.

\begin{figure}[b]
    \centering
    \includegraphics[width=0.9\columnwidth]{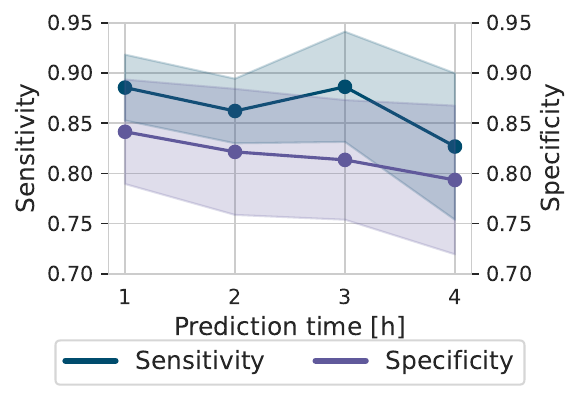}
   \caption{Retrospective analysis over different prediction times.}
   \label{fig:offline_results}
\end{figure}

\begin{figure*}
    \centering
    \includegraphics[width=\textwidth]{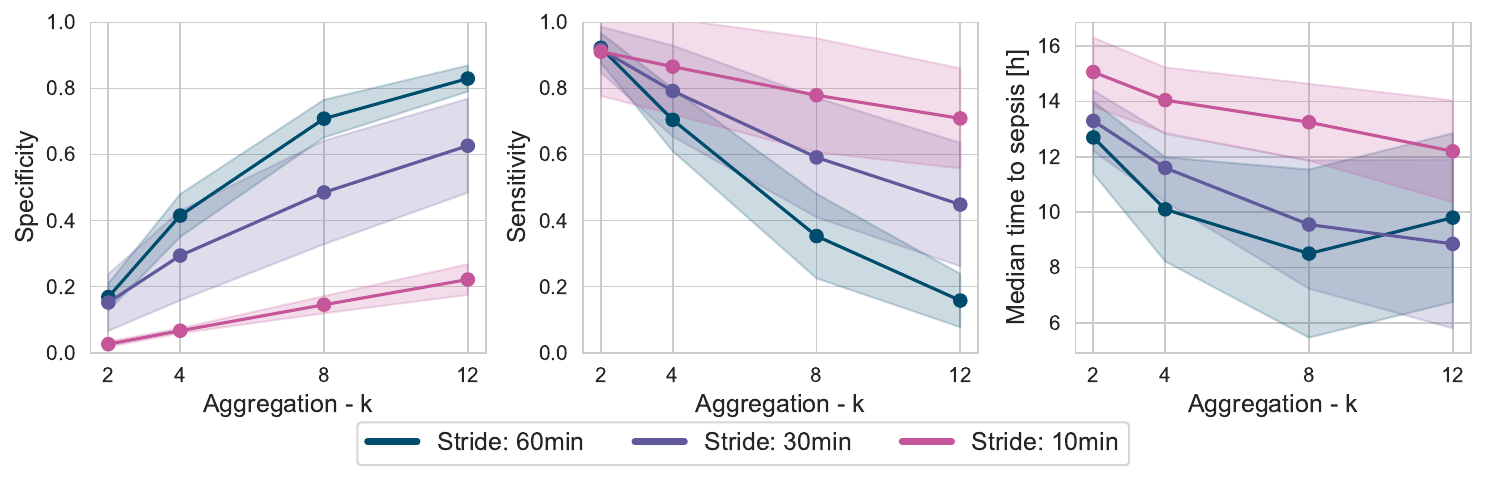}
   \caption{Parameter study on the online model varying the windows strides and the aggregation parameter.}
   \label{fig:online_results}
\end{figure*}

\subsection{Model quantisation}
The nRF5340 from Nordic Semiconductor was used for model deployment. The MCU features a Cortex M33, which \rebu{has been clocked at \qty{128}{\mega\hertz}} and provides a RAM of \qty{512}{\kilo\byte} and a flash of \qty{1}{\mega\byte}. Hence, the TCN model weights and activation parameters were quantized to an 8-bit integer. The CMSIS-NN\cite{lai2018cmsis} library was used to accelerate inference on the MCU on both the 1D convolutions, including the dilated kernels, and the dense layers while exploiting single instruction multiple data (SIMD) instruction in the Cortex-M33 instruction set. Tensorflow Light for MCU\cite{david2021tensorflow} was chosen as a framework to port the neural network on hardware. The quantization was performed after training, and to optimize the dynamic range a representative dataset was passed to the network. The samples were taken randomly from the train set to offer representative enough data, while not introducing any unwanted spillage between the train and the test datasets. Additionally, power characterization of the deployment algorithms was conducted to evaluate the feasibility of real-world operation. 
\begin{table}[t]
\centering
\begin{tabular}{lcc}
Metric      & Float & INT8 \\
\hline
Sensitivity  & 0.59          & 0.60   \\
Specificity  & 0.49           & 0.48  \\
Med. t. sep. & \qty{9.6}{\hour}          & \qty{9.8}{\hour}   \\
\end{tabular}
\caption{Performance comparison between floating point and quantized model.}
\label{tab:quant_float_comp}
\end{table}

\begin{table}[t]
\centering
\begin{tabular}{lcc}
Metric      & Float & INT8 \\
\hline
MAC         & 2e6   & 2e6  \\
Weights     & \qty{412}{\kilo\byte}  & \qty{103}{\kilo\byte} \\
Activations & \qty{92}{\kilo\byte}  & \qty{23}{\kilo\byte}  \\
MAC/Cycle   & -     & 0.11  \\
Latency     & -     & \qty{143}{\milli\second}  \\
Energy      & -     & \qty{2.68}{\milli\joule} \\
\end{tabular}
\caption{Neural network evaluation metrics, implemented on the MCU.}
\label{tab:nn_prof}
\end{table}

\section{Results and discussion}

\subsection{Deep Learning}
\label{sec:res_dl}
\subsubsection{Model baseline: Retrospective sepsis identification}
A retrospectively trained TCN model with only vital signs showed a sensitivity of 0.83 and specificity of 0.79 at a prediction horizon of 4 hours to sepsis, as shown in Figure \ref{fig:offline_results}. This shows that it is possible to identify features of sepsis using only vital sign data that can be digitally acquired from a wearable system. However, this model by itself cannot be used for real-time implementation as it assumes knowledge of where sepsis occurs in time and this is unknown in a realistic condition.

\subsubsection{Real-time sepsis onset identification}

Figure \ref{fig:online_results} shows the specificity, sensitivity, and median time to sepsis as predicted by the model for different values of aggregation parameter \textsc{k} for three values of stride of 10 minutes, 30 minutes, and 60 minutes. The solid line reports the mean and the shaded area represents the standard deviation over a 5-fold cross-validation done on the test dataset. As observed in Figure \ref{fig:online_results}, a shorter window stride exhibits a better true negative rate than longer window strides but has a lower true positive rate. This is possible due to the high overlap between two consecutive windows moved by a shorter stride which introduces redundancy while a larger window might miss the causal relationship between the sepsis informative features resulting in lower specificity and greater false positive rate.  

The specificity and sensitivity metrics show a proportional and inverse proportional relationship to the aggregation parameter \textsc{k} respectively. Increasing the aggregation parameter \textsc{k} leads to tuning the consensus algorithm to have the prediction model more reliant on consecutive positively predicted windows, hence having higher specificity. However, this very requirement would necessitate more positive predictions to raise an alarm thereby increasing the rate of false negatives.
Hence, for calculating the algorithm's performance in terms of median time to sepsis, a value of aggregation parameter \textsc{k} and the window stride was chosen that showed a good trade-off between the sensitivity and specificity. This was set to be a stride of 30 minutes and an aggregation parameter \textsc{k} of 8, which had a sensitivity of (0.59 $\pm$  0.18) and specificity of (0.49 $\pm$  0.16). Using these parameters, the median predicted time to sepsis was found to be \qty{9.6}{\hour} (CI: [\qty{7.3}{\hour}, \qty{11.9}{\hour}]) before sepsis onset.

The real-time implementation shows performance degradation when compared to the retrospective sepsis prediction. One reason could be labeling all the windows of a sepsis-positive patient as class 1, which was based on the assumption that the features leading to the adverse event might be present from the beginning. This might not be the case for all patients, in reality, leading to errors in training. Moreover, this evaluation is limited to the HiRID dataset which is not representative of a large sample diversity. However, we were limited by this choice of the dataset as it is the only open-source data set with high granularity vital sign measurement.

\subsection{Quantisation}
Table \ref{tab:quant_float_comp} shows a comparison of the floating point and quantized model for the chosen model. The performance of the floating point and integer models are very close, within 2\% for all three metrics, sensitivity, specificity, and median time to sepsis. In particular, the quantized model shows a slightly better performance than the floating point model in terms of sensitivity (0.60 vs 0.59) and median time to Sepsis (9.8 vs 9.6) hours and a small degradation in specificity (0.48 vs 0.49).

The inference profiling is reported in Table \ref{tab:nn_prof}. The network accounts for only 2 million multiply-accumulate (MAC) operations, which results in a latency of \qty{143}{\milli\second} on the target architecture, therefore yielding a throughput of 0.11 MAC/Cycle. Weights and activations of the quantized model require \qty{103}{\kilo\byte} and \qty{23}{\kilo\byte} respectively, reducing the memory footprint by a factor 4x from the float point arithmetic the model was trained on. Runtime results for the floating model are not reported since the model could not be run on the MCU, given the overhead of the Zephyr RTOS and the TensorflowLight for Micro runtime used for the deployment.

\section{Conclusions}

This work presented a lightweight quantized neural network for sepsis detection that can be deployed in wearable devices and uses only 6 vital signs extracted by low-power sensors such as PPG, IMU and body temperature sensors. Targeting low-power sensors, SepAl achieves real-time early detection of sepsis with a median predicted time to sepsis of 9.8 hours before onset, with a sensitivity and specificity of 0.60 and 0.48, respectively. 
The neural network proposed in this work is fully quantized and deployable in mW-range Bluetooth low-energy microcontrollers, enabling
\rebu{
future research in compact and low-power hardware to integrate sensing, vital sign extraction, and inference in one single device.}
Through the utilization of digital biomarkers and a temporal convolutional neural network, our system enables real-time analysis, timely alerts, and proactive intervention. SepAl represents a promising advancement in wearable healthcare technology, with the potential for improving patient outcomes and revolutionizing sepsis management. 

\section*{Acknowledgment}
This work was supported by the Basel Research Center for Child Health (BRCCH), Switzerland.

\bibliographystyle{ieeetr}
\bibliography{references.bib}

\end{document}